\documentstyle[12pt]{article}
\def\be{\begin{equation}}
\def\ee{\end{equation}}
\def\ba{\begin{array}}
\def\ea{\end{array}}
\def\bea{\begin{eqnarray}}
\def\eea{\end{eqnarray}}
\def\br{\begin{eqnarray}}
\def\er{\end{eqnarray}}

\begin{document}
\begin{center}
{\Large \bf Shell closure effects studied via cluster decay in heavy nuclei} \\
\bigskip
{\bf Sushil Kumar$^{1}$, Ramna $^{1}$, Rajesh Kumar$^{2}$}\\

\medskip

$^1$ Department of Physics, Chitkara University, solan-174103, H.P., India\\
$^2$ Department of Applied Sciences and Humanities, N.I.T. Hamirpur-177005, H.P., India.\\

\medskip
\end{center}
\vspace*{0.3cm}
\begin{center}
{\bf Abstract}
\end{center}
\medskip
\baselineskip 18pt

The effects of shell closure in nuclei via the cluster decay is
studied. In this context, we have made use of the Preformed
Cluster Model ($PCM$) of Gupta  and collaborators based on the Quantum
Mechanical Fragmentation Theory. The key point in the cluster
radioactivity is that it involves the interplay of close shell effects of
parent and daughter. Small half life for a parent indicates shell stabilized
daughter and long half life indicates the stability of the parent against the decay.
In the cluster decay of trans lead nuclei observed so far, the end product is doubly magic lead or its neighbors. With this in our mind we have extended the idea of cluster radioactivity. We investigated decay of different nuclei where Zirconium is always taken as a daughter nucleus, which is very well known deformed nucleus. The branching ratio of cluster decay and $\alpha$-decay is also studied for various nuclei, leading to magic or almost doubly magic daughter nuclei. The calculated cluster decay half-life are in well agreement with the observed data. First time a possibility of cluster decay in $^{218}U$
nucleus is predicted.

\vfill\eject
\section {\large \bf Introduction}
When a charged particle heavier than $^4_2He$, but lighter than a
fission fragment is emitted by an unstable nucleus, the process is
called cluster radioactivity or heavy ion radioactivity. Cluster radioactivity was predicted theoretically in $1980$ by S\v andulescu, Poenaru and Greiner \cite{psandu80} on the basis of Quantum Mechanical Fragmentation Theory ($QMFT$). The first
experimental identification of a case of radioactive decay of
heavy nuclei by emission of a nuclear fragment heavier than
$^4_2He$ was done by Rose and Jones \cite{prose84}, who studied
the decay of $^{223}Ra$ by $^{14}C$ emission with a half-life
of $(3.7\pm1.1)\times10^7$ yr.
Rose and Jones measured a branching ratio of $B(^{14}C/\alpha) =
8.5(2.3)\times 10 ^{-10}$ . This result was then confirmed by
Alekshandrov \textit{et al.} \cite{Aleksandrov84}, Gale \textit{et
al.} \cite{Gale84} and Price \textit{et al.} \cite{Price85}.\\

Cluster radioactivity has grown and there are more than $24$ cases
of cluster radioactivity with partial
half-lives ranging from $log_{10}T$ = $>3.63(s)$ to $>29.20(s)$
\cite{bonetti07}. There exists a whole family of cluster decay
modes like $^{14}C$ radioactivity, $^{24}Ne$ radioactivity,
$^{28}Mg$ radioactivity and so on. The mother nuclei range from
$^{221}Fr$ to $^{242}Cm$, all from trans-lead region and cluster
radioactivity in this region indicates the presence of heavier
clusters. Shell effects are clearly manifested in cluster radioactivity since
experiments show that the cluster decay result in magic daughter
products.\\

The study of $\alpha$-decay and cluster decay has been used for identifying the
shell closure effects including even the very weak sub-shell closures \cite{sushil02, sushil03,sharda03, gupta91}. Both the cases of large and small decay half-lives are important. Small ones imply closed shell effects in daughter nucleus and the large ones indicate closed shell effects of the mother nucleus. In decay calculations presence of the known spherical or deformed daughter should result in a small decay half-life or a small decay half-life should refer to the existence of a known or new, spherical or deformed, closed shell for the daughter nucleus.\\

Superdeformation in medium-mass nuclei is a subject of interest and it
was first discovered in the actinide fission
isomers \cite{ppolik62}. Later it was explained from the secondary
minimum at very large deformation\cite{pstrut68}.
$\gamma$- ray transtions between states in the $N=Z$ nucleus $^{80}Zr$ was studied to
see the deformation, and it was observed that this is the most deformed nucleus in nature, with
a quadrupole deformation $\beta_2$ $\approx 0.4$ \cite{plister87}.
Superdeformation band in the $N,Z=40$ mass region has also been studied experimentally
by the Baktash et.al., \cite{baktash95} and a significant subshell closure at the
$N=40$ neutron number has been studied in the $^{68}Ni$ \cite{broda95}.\\

We have not performed the cluster decay calculations, only to search the
spherical and/or deformed magicity i.e. not to just study the
stability /instability of the concerned nuclei, but also for looking
the most probable cluster decay modes of $^{218}$U nucleus that has recently
been studied for  $\alpha$-decay \cite{plepp07}. \\

We are using the Preformed Cluster Model(PCM) of Gupta and Collaborators, where the cluster
is assumed to be preformed in the mother nucleus. In this model the preformation probability (also known as spectroscopic factor in various models) for different possible clusters are calculated by solving the Schrodinger equation for the dynamical flow of mass and charge.Another such model based on the calculations of spectroscopic factors for different clusters, has also been given Blendowske and Walliser \cite{pblende87,pblende88}. In this model they have given parameterization of the spectroscopic factor and meanwhile, other contributions have extended these results \cite{hess04}. Also,  the so called " Semimicroscopic Algebraic Cluster Model" \cite{cseh92,cseh94,cseh91,pchess04} contributes to parameterization of the spectroscopic factors. Deformations of the neutron rich clusters have effect on penetration probability as the inclusion of the deformations reduces the barrier height. This fact is incorporated into $PCM$ through parameter $R$, which results in the lowering of barrier height\cite{pkumar97}.
Recently \cite{csehc04,algora06}, role of deformation on binary and ternary clusterization has been studied to investigate the exotic nuclear shapes, on the basis of U(3)selection rule .\\

The paper is organised as follows. The calculations
are made by using the preformed cluster-decay model
(PCM) of Gupta and collaborators \cite{pkumar97,prkg88,pmalik89} whose
outline is presented in section II. Section III deals with
the calculations and results obtained from this study.

\section {\large \bf The Preformed Cluster Model (PCM)}

The Preformed Cluster Decay Model (PCM)\cite{pkumar97,prkg88,pmalik89} is based on the Quantum Mechanical Fragmentation Theory \cite{gupta99,maruhn74,gupta75}. The dynamical collective coordinates of mass $\eta$ and charge asymmetries $\eta _Z$ are in addition to the usual coordinates of relative separation R.

The decay constant $\lambda $ (or the decay half-life $T_{1/2}$ ) in PCM is defined as
\be \lambda ={{{ln 2}\over {T_{1/2}}}}=P_0\nu _0 P. \label{eq:1}
\ee Here $P_0$ is the cluster (and daughter) preformation
probability and P the barrier penetrability which refer,
respectively, to the $\eta$ and R motions. The $\nu _0$ is the
barrier assault frequency. The $P_0$ are the solutions of the
stationary Schr\"odinger equation in $\eta$, \be \{
-{{\hbar^2}\over {2\sqrt B_{\eta \eta}}}{\partial \over {\partial
\eta}}{1\over {\sqrt B_{\eta \eta}}}{\partial\over {\partial \eta
}}+V_R(\eta )\} \psi ^{({\nu})}(\eta ) = E^{({\nu})} \psi
^{({\nu})}(\eta ), \label{eq:2} \ee which on proper normalization
are given as \be P_0={\sqrt {B_{\eta \eta}}}\mid \psi
^{({0})}(\eta (A_i))\mid ^2\left ({2/A}\right ), \label{eq:3} \ee
with i=1 or 2 and $\nu$=0,1,2,3.... Eq. (\ref{eq:2}) is solved at a
fixed $R=R_a=C_t(=C_1+C_2)$, the first turning point in the WKB
integral for penetrability P (Eq. \ref{eq:5}).
The $C_i$ are S\"ussmann central radii
$C_i=R_i-({1/R_i})$, with the radii
$R_i=1.28A_i^{1/3}-0.76+0.8A_i^{-1/3} fm$.\\

The fragmentation potential $V_R(\eta )$ in (\ref{eq:2}) is
calculated simply as the sum of the Coulomb interaction, the
nuclear proximity potential \cite{blocki77} and the ground state
binding energies of two nuclei, \be V(R_a, \eta) =\sum_{i=1}^{2} B(A_{i}, Z_{i})
+ \frac{Z_{1} Z_{2} e^{2}}{R_a} + V_{P}, \label{eq:4} \ee with B's taken from the 2003
experimental compilation of Audi et al., \cite{audi03} and from
the 1995 calculations of M\"oller et al. \cite{moller95} whenever
not available in \cite{audi03}. Thus, full shell effects are
contained in our calculations that come from the experimental
binding energies and/or from the calculations of M\"oller et al.
\cite{moller95}. The charges Z$_1$ and Z$_2$ in (\ref{eq:4}) are
fixed by minimizing the potential in $\eta_Z$ coordinate with Z.
The Coulomb and proximity potentials in (\ref{eq:4}) are for spherical
nuclei. The mass parameters $B_{\eta \eta}(\eta )$, representing
the kinetic energy part in (\ref{eq:2}), are the classical
hydrodynamical masses \cite{kroger80}.\\

The WKB tunnelling probability is
$P=P_i P_b$ with \be P_i=exp[-{2\over
\hbar}{{\int }_{R_a}^{R_i}\{ 2\mu [V(R)-V(R_i)]\} ^{1/2} dR}]
\label{eq:5}
\ee \be P_b=exp[-{2\over \hbar}{{\int }_{R_i}^{R_b}\{
2\mu [V(R)-Q]\} ^{1/2} dR}]. \label{eq:6} \ee
These integrals are solved analytically \cite{pmalik89} for $R_b$, the second turning
point, defined by $V(R_b)=Q$-value for the ground-state decay.

The assault frequency $\nu _0$ in (\ref{eq:1}) is given simply as
\be \nu _0=(2E_2/\mu )^{1/2}/R_0, \label{eq:7} \ee with
$E_2=(A_1/A) Q$, the kinetic energy of the lighter fragment, for
the $Q$-value shared between the two products as inverse of their
masses. Here $R_{0}$ is the equivalent spherical radius of the mother nucleus\\

\section {\large \bf Results and Discussion}

In this paper we have performed three types of calculations.
Firstly, we have focused on the search for the
magicity (spherical and /or deformed) taking the Zirconium nucleus as
daughter in the decay of mother nuclei, which are taken from mass region $A=88$ to $A=186$. Secondly, we have investigated the cluster
decay of various nuclei taken from the heavy mass region  and have also compared data
with experimental results. Thirdly, we have searched for the possible cluster decay modes of $^{218}U$ nucleus, whose alpha decay has been studied recently \cite{plepp07}, along with few other isotopes of Uranium ($^{230,232,234,236}U$).\\

In the first part of our calculation we studied the decay of different nuclei taken from the intermediate mass region. We considered the Zirconium nucleus as daughter product in the decay of all these nuclei. Mother of even
mass number is chosen and it results in a daughter with different even neutron number. The decaying nuclei are from $^{88}Ru$,
$^{86-90}Pd$,$^{90-96}Cd$,$^{100-108}Cd$, $^{94-116}Sn$,
$^{100-126}Te$, $^{158-166}Te$,$^{104-134}Xe$, $^{148-176}Xe$ to
$^{108-186}Ba$. Selection of isotopes of different mother nuclei for the
calculations has been made on the basis of Q value and  the decay of only those was considered for which it is positive.\\


The histogram in $Fig.(1)$, contains maxima and minima of half-lives of above mentioned parent nuclei with neutron number of the daughter. The daughter was fixed to be Zirconium nucleus i.e. $Z_{1}=40$. We considered all possible cluster decay channels of above mentioned nuclei. Half lives for different decaying parents giving $Z_{1}=40$ and corresponding neutron number $N_{1}$ of the daughter with the complementing cluster were calculated. There exists different combinations of parents and clusters giving rise to $Z_{1}=40$ with a particular value of $N_{1}$ i.e. a particular isotope of Zirconium, as a daughter product. Among different values of  half lives of all such possible decays resulting into the given Zirconium isotope, we noted the maximum and minimum values only. Then these two extreme values of the half life were plotted for different isotopes of Zirconium. \\

In another histogram the $Fig.(2)$, in which even neutron number of daughter nuclei vs
half-lives of various mother nuclei are shown. The height of the
bar corresponding to the daughter neutron number represents the
stability/ instability of the daughter nucleus. In this
calculation the shell closure effects appear at $N_{1}=40$ and at
$N_{1}=82$. Also at $N_{1}=70$, these appear, but weakly. \\

Shell effects can be seen in $Fig.(3)$ also, where we have
displayed half-lives for the following decay modes: $^{14}C$,
$^{18-20}O$, $^{24}Ne$, versus the neutron number of the daughter,
$N_{d}$. The half-lives shows the minima at the magic number for
all the decay mode considered above. The $Q$-value is one of the physical quantity which plays a very
important role in any spontaneous nuclear decay with emission of
charged particles. The $Q$-value
behavior has been noticed for these calculations, as the size
of the cluster increases $Q$-value also increases \cite{sushil02}.
The calculations belonging to the same atomic number of the daughter
are joined with a line of a style mentioned in the figure for
$Z_{d} = 82$ and $84$. For the $^{14}C$ radioactivity where daughter
is always with same atomic number and neutron numbers vary, we
have observed in our calculations that $N_{d}=126$ appears as a magic
number in all the cases, same is for the $^{20}O$ cluster when the
daughters are $ Z_{d}= 80, 82$.  A strong shell effect can be seen
in this figure. As a rule, the shortest value of the half-life is
obtained when the daughter nucleus has a magic number of neutrons
$N_{d} =126$ and protons $Z_{d}=82$.\\

Same calculation has been made but for different clusters taking
the same atomic number of different daughter nuclei $Z_{d}=
80,81,82,83,84$ . For the $^{18}O$ radioactivity where daughter is
always with same atomic number and neutron number vary, we have
observed that $N_{d}=126$ appears as a magic
number in all the cases, same is for the $^{24}Ne$ cluster when
the daughter are $Z_{d}= 80,81,82$. Again a strong shell effect
can be seen in $Fig.3$. The most probable cluster from heavy nuclei
in $PCM$ model and their characteristics (i.e., Q-value and
Half-life) are compared with the experimental data (given in
\cite{bonetti07,poenaru96,poenaru02,poenaru06} and the references therein) in Table 1.\\

In the third part of the calculations, we have predicted for the first time
the possible cluster decay modes of $^{218}U$ which was recently investigated for alpha-decay \cite{plepp07}.
The $\alpha$-decay calculation of $^{218}U$ in the
Quantum Mechanical Fragmentation Theory has been studied and we
have calculated the decay properties not only for $\alpha$-decay
but also for other possible clusters. The half life for for $\alpha$-decay, using PCM
calculations comes out to be $T_{1/2}=4.36ms$ and its experimental value is $T_{1/2}=0.51ms$.
The calculated Q-value is, $E_{\alpha}^{PCM}=8.788MeV$ and its experimental value is $E_{\alpha}^{exp}=8.612MeV$.
 Shell closure effects play an important role in the
cluster decay studies. Predicted half-lives $T_{1/2}$(s) and other
characteristics for this isotope of Uranium are given in table(2). $^{8}Be$, $^{12}C$ and $^{22-24}Ne$ appear as the most probable cluster decay modes in our calculations.\\

\begin{table}
\caption{Comparision of experimental $Q$-values, Half-lives with
calculated $Q$-values, Half-lives in PCM $\alpha$-decay of cluster
emitters.}
\begin{center}
\begin{tabular}{|l|c|c|c|c|c|c|} \hline
Parent nucleus&Emitted cluster & $Q^{exp.}$ & $Q^{PCM.}$  & $T_{1/2}^{exp}$& $T_{1/2}^{PCM}$& \\
Z and A   & $Z_{e}$ and $A_{e}$& MeV      & MeV       & sec.         & sec           & \\
          &                    &          &           &              &               & \\ \hline
87 - 221 & 6- 14 & 31.294& 31.292 &14.53&18.044 &\\
88 - 221 & 6 - 14&32.402 & 32.395   &13.39& 17.25       &\\
88- 222  & 6- 14 &33.052 & 33.05    &11.01& 15.098      &\\
88- 223  & 6 -14 &31.839 & 31.829    &15.15& 18.548       &\\
88- 224  & 6- 14 &30.541 & 30.536    &15.69& 19.676       &\\
88 -226  & 6- 14 & 28.198& 28.197    & 21.22& 25.805      &\\
89- 225  &6 -14  & 30.479& 30.477     & 17.16&  22.664    &\\
90- 226  & 8 -18 & 45.731 & 45.726    & $>$16.76& 23.289     &\\
90- 228 & 8- 20  &44.730  & 44.724    &20.72 & 23.357     &\\
90- 230 & 10- 24 &57.765  &  57.759    &24.61 & 26.849     &\\
90 - 232 &10 - 24& 54.491 &      &$>$29.20&      &\\
90 - 232 &10 - 26& 55.973 & 55.964     &$>$29.20 & 29.831    &\\
91- 231& 10 -24& 60.413   & 60.409      &22.88 & 24.999     &\\
91 -231 & 9 - 23& 51.854  & 51.844      &$>$26.02 & 26.463     &\\
92- 230 &10- 22 &  61.390 &      &19.57 &        &\\
92- 232 &10- 24 &62.312   & 62.309     & 20.42 & 21.528       &\\
92 -233 &10- 25 &60.736 &        & 24.84 &        &\\
92- 233 & 10- 24 & 60.490 &       & 24.84 &       &\\
92 -233 & 12- 28 &74.235&        &27.59 &          &\\
92- 234& 12- 28 &74.118 &74.109        &25.74 & 27.746        &\\
92- 234 & 10- 24 &58.831& 58.825       &$>$25.92 & 28.775       &\\
92 -234 & 10- 26 &59.473& 59.465       &25.92 & 26.411      &\\
92- 235 &10 -24  & 57.358 &57.362       &27.42 & 32.511        &\\
92- 235 &10 -25 &57.717 &  57.756       &27.42 &31.173        &\\
92- 235 &12- 28 &72.162&         &$>$28.09 &         &\\
92 - 236&  12- 28& 70.558&       & 27.58 &          &\\
92- 236  &12- 30& 72.280&        & 27.58 &        &\\
93-237   &12-30 & 74.791 &  74.816     &$>$27:57 & 28.524      &\\
94- 236 & 12-28& 79.674 & 79.668       &21.67 & 21.637       &\\
94- 238 &14- 32& 91.198& 91.189        & 25.27& 27.069        &\\
94- 238 &12- 28 &75.919& 75.91         &25.70& 28.395        &\\
94- 238 &12- 30 &76.801& 76.822        & 25.70& 25.542       &\\
94- 240 &14- 34 &91.038& 91.027        &$>$25:52 & 25.904       &\\
95- 241 &14- 34 &93.931& 93.925        &$>$25:26 & 23.204       &\\
96- 242 &14- 34 &96.519& 96.509        & 23.15&  21.08      &\\
\hline
\end{tabular}\\[0.5ex]
{}
\end{center}
\end{table}

\begin{table}
\caption{Predicted half-lives $T_{1/2}$(s) and other
characteristics for cluster decays of Uranium nuclei. For the
Q-value estimates the masses for these parents are taken from the
Audi et. al. mass table \cite{audi03} and
Moller- Nix \cite{moller95}.}\label{f:ct1}

\begin{center}
\begin{tabular}{|c|c|c|c|c|c|c|c|}
\hline
Parent &Emitted &Daughter & Log $T_{1/2}$ & Preformation   & Decay   & Q value & \\
nucleus&cluster &nucleus  & sec       & probability    &constant & MeV      &  \\
       &        &         &           & P$_{o}$      &                  &   & \\
\hline

$^{218}$U & $^{4}$He & $^{214}$Th & -2.36 & 2.50$\times 10^{-05}$&
1.59$\times10^{02}$& 8.788&\\
&$^{8}$Be & $^{210}$Ra & $19.759$ & 9.72$\times$ 10$^{-16}$&
1.21$\times$10$^{-20}$& 16.521&\\
& $^{10}$Be & $^{208}$Ra & $74.199$ & 4.68$\times$ 10$^{-21}$&
4.38$\times$10$^{-75}$& $7.604$&\\
&$^{12}$C & $^{206}$Rn & 24.992 & 1.07$\times$ 10$^{-22}$&
7.06$\times$10$^{-26}$& 31.037&\\
& $^{18}$O & $^{200}$Po & 41.371 & 7.68$\times$ 10$^{-30}$&
2.95$\times$10$^{-42}$& 39.658&\\
& $^{20}$O & $^{198}$Po & 58.519 & 4.09$\times $10$^{-33}$&
2.10$\times$10$^{-59}$& 33.6&\\
& $^{22}$Ne & $^{196}$Pb & 39.92 & 2.99$\times$ 10$^{-32}$&
8.33$\times$10$^{-41}$& 55.307&\\
& $^{24}$Ne & $^{194}$Pb & 46.19 & 2.83$\times$ 10$^{-33}$&
4.48$\times$10$^{-47}$& 52.081&\\
& $^{28}$Mg & $^{190}$Hg & 42.169 & 3.85$\times$ 10$^{-34}$&
4.70$\times$10$^{-43}$& 68.311&\\
\hline
$^{230}$U & $^{4}$He & $^{226}$Th & 8.024 & 1.15$\times$
10$^{-08}$&6.56$\times$ 10$^{-09}$& 5.993&\\
&$ ^{10}$Be & $^{220}$Ra & 64.1 & 8.55$\times$ 10$^{-22}$&
5.50$\times$10$^{-65}$& 8.736&\\
& $^{14}$C & $^{216}$Rn & 30.109 & 3.04$\times$ 10$^{-22}$&
5.40$\times$10$^{-31}$& 28.34&\\
& $^{20}$O &$^{210}$Th & 28.69 & 2.06$\times$ 10$^{-23}$&
1.42$\times$10$^{-29}$& 43.771&\\
& $^{23}$F & $^{207}$Bi & 34.014 & 1.48$\times$ 10$^{-25}$&
6.71$\times$10$^{-35}$& 48.339&\\
& $^{24}$Ne & $^{206}$Pb & 24.286 & 7.45$\times$ 10$^{-24}$&
3.59$\times$10$^{-25}$& 61.35&\\
& $^{26}$Ne & $^{204}$Pb & 33.291 & 1.62$\times$ 10$^{-26}$&
3.55$\times$10$^{-34}$& 56.294&\\
\hline
$^{232}$U& $^{4}$He & $^{228}$Th & 10.913 & 5.41$\times$
10$^{-09}$&8.47$\times$ 10$^{-12}$& 5.414&\\
& $^{10}$Be & $^{222}$Ra & 74.01 & 4.14$\times$ 10$^{-22}$&
6.77$\times$10$^{-75}$& 7.683&\\
& $^{14}$C & $^{218}$Rn & 35.302 & 2.41$\times$ 10$^{-23}$&
3.46$\times$10$^{-36}$& 26.374&\\
& $^{22}$O &$ ^{210}$Po & 32.841 & 5.05$\times$ 10$^{-23}$&
9.99$\times$10$^{-34}$& 41.279&\\
& $^{24}$Ne & $^{208}$Pb & 21.528 & 3.78$\times$ 10$^{-22}$&
2.06$\times$10$^{-22}$& 62.309&\\
& $^{26}$Ne & $^{206}$Pb & 28.832 & 2.98$\times$ 10$^{-24}$&
1.02$\times$10$^{-29}$& 57.966&\\ \hline
\end{tabular}
\end{center}
\end{table}
\begin{table}{Table \ref{f:ct1}: Continued.....} \\ \\
\begin{center}
\begin{tabular}{|c|c|c|c|c|c|c|c|} \hline
Parent &Emitted &Daughter & Log $T_{1/2}$ & Preformation   & Decay   & Q value & \\
nucleus&cluster &nucleus  & sec       & probability    &constant & MeV      &  \\
       &        &         &           & P$_{o}$               &         &          & \\ \hline
$^{234}$U& $^{4}$He & $^{230}$Th & $14.443$ & 1.64$\times
$10$^{-09}$& 2.50$\times$ 10$^{-15}$& $4.859$&\\
&$^{10}$Be & $^{224}$Ra & 86.579 & 1.04$\times$ 10$^{-23}$&
1.83$\times$ 10$^{-87}$& 6.713&\\
&$^{14}$C & $^{220}$Rn & 42.214 & 9.38$\times$ 10$^{-26}$&
4.24$\times$10$^{-43}$& 24.514&\\
& $^{22}$O & $^{212}$Po & 38.903 & 1.35$\times$ 10$^{-25}$&
8.66$\times$10$^{-40}$& 39.231&\\
& $^{24}$Ne & $^{210}$Pb & 28.775 & 7.96$\times$ 10$^{-26}$&
1.16$\times$ 10$^{-29}$& 58.825&\\
& $^{26}$Ne & $^{208}$Pb & 26.411 & 1.14$\times$ 10$^{-23}$&
2.69$\times$ 10$^{-27}$& 59.465&\\
\hline
$^{236}$U& $^{4}$He & $^{232}$Th & $16.521$ & 8.58$\times$
10$^{-10}$& 2.09$\times$ 10$^{-17}$& $4.574$&\\
& $^{10}$Be & $^{226}$Ra & $93.871$ & 1.06$\times$
10$^{-23}$& 9.33$\times$ 10$^{-95}$& $6.171$&\\
& $^{14}$C & $^{222}$Rn & $47.102$ & 1.57$\times$
10$^{-26}$& 6.74$\times$ 10$^{-48}$& $23.054$&\\
& $^{22}$O & $^{214}$Po & $42.807$ & 1.77$\times$
$10^{-26}$& 1.08$\times$ $10^{-43}$& $37.631$&\\
& $^{24}$Ne & $^{212}$Pb & $34.18$ & 8.13$\times$
$10^{-28}$& 4.58$\times$ $10^{-35}$& 55.944  &\\
& $^{26}$Ne & $^{210}$Pb & $31.417$ & 1.50$\times$
10$^{-25}$& 2.65$\times$ 10$^{-32}$& $56.745$&\\ \hline
\end{tabular}\\[0.5ex]
{}
\end{center}
\end{table}

\newpage
\par\noindent
{\bf Figure Captions}\\

Fig.1:Predicted half-lives minimum and maximum with daughter (Zr)
neutron number for various cluster decay mode.

Fig.2: Predicted half-lives (minimum only) plotted as a function
of the daughter neutron number for various cluster decay mode.

Fig.3:Predicted half-lives for the following decay modes:
$^{14}$C, $^{18-20}$O, $^{24}$Ne, versus the neutron number of the
daughter, $N_{d}$.


\newpage
\begin{thebibliography}{999}
\bibitem{psandu80}
Sandulescu A., Poenaru D. N. and Greiner W. 1980 Sov. J. Part. Nucl. {\bf 11}
p. 528.
\bibitem{prose84}
Rose H. J. and Jones G. A. 1984 Nature {\bf307} p. 245.
\bibitem{Aleksandrov84}
Aleksandrov D. V. et al. 1984 Pis'ma Zh. Eksp. Teor. Fiz. {\bf40} 152.
\bibitem{Gale84}
Gale S. et al.1984 Phys. Rev. Lett. {\bf53} 759.
\bibitem{Price85}
Price P.B. et al. 1985 Phys. Rev. Lett. {\bf54} 297.
\bibitem{bonetti07}
Bonetti R., Guglielmetti A. 2007 Romanian Reports in Physics Vol. {\bf59} No. 2 P. 301-310.
\bibitem{sushil02}
Gupta R. K., Sushil Kumar, Rajesh Kumar, Balasubramaniam M. and Scheid W.
2002 J. Phys. G: Nucl. Part. Phys. {\bf28} 2875-2884.
\bibitem{sushil03}
Sushil Kumar, Balasubramaniam M.,Gupta R. K., M\"unzenberg and Scheid W. 2003
J. Phys. G: Nucl. Part. Phys. {\bf29} 625-639.
\bibitem{sharda03}
Gupta R. K. et al., 2003 Phys. Rev. C{\bf 68} 034321.
\bibitem{gupta91}
Gupta R. K., Scheid W. and Greiner W, 1991 J.Phys.G: Nucl. Part. Phys. {\bf 17}
1731.
\bibitem{ppolik62}
Polikanov S. M. et al., 1962 Sov. Phys. JETP {\bf 15} 1016.
\bibitem{pstrut68}
Strutinsky V. M., 1967 Nucl. Phys. {\bf A95} 420 ; 1968 {\bf A122} 1.
\bibitem{plister87}
Lister C. J. et al., 1987 Phys. Rev. Lett. {\bf 59} 1270.
\bibitem{baktash95}
Baktash C. et al., 1995 Phys. Rev. Lett. {\bf 74} 1946.
\bibitem{broda95}
Broda R. et al., 1995 Phys. Rev. Lett. {\bf 74} 868.
\bibitem{plepp07}
Leppanen A. P. et al., 2007 Phys. Rev. {\bf C75} 054307.
\bibitem{pblende87}
Blendowske R., Fliessbach T., Walliser H., 1987 Nucl.Phys.{\bf A464}75.
\bibitem{pblende88}
Blendowske R. and Walliser H., 1988 Phys. Rev. Lett. {\bf 61} 1930.
\bibitem{hess04}
Hess P. O., Misicu S., 2004 Phys. Lett. {\bf B 595} 187-192.
\bibitem{cseh92}
Cseh J., 1992 Phys. Lett. {\bf B 281} 173-177.
\bibitem{cseh94}
Cseh J. and Levai, 1994 Annals of Physics {\bf 230} 165-200.
\bibitem{cseh91}
Cseh J. et al., 1991 Phys. Rev. {\bf C 43} 165.
\bibitem{pchess04}
Hess P.O. et al., 2004 Phys. Rev. {\bf C 70} 051303(R).
\bibitem{pkumar97}
S. Kumar and R. K. Gupta, 1997 Phys. Rev. {\bf C55} 218.
\bibitem{csehc04}
Cseh J. et al., 2004 Phys. Rev. {\bf C 70} 034311.
\bibitem{algora06}
Algora A. et al., 2006 Phys. Lett. {\bf B 639} 451-455.
\bibitem{prkg88}
Gupta, R. K. 1988, Proc. 5th Int. Conf. Nuclear Reaction
Mechanisms (Varenna) ed E Gadioli (Milano:Ricerca Scientifica ed
Educazione Permanente) Giessen Preprint.
\bibitem{pmalik89}
Malik S. S. and Gupta R. K., 1989 Phys. Rev. {\bf C39} 1992.
\bibitem{gupta99}
Gupta R. K. and Greiner W., 1999 Heavy Elements and Related New Phenomena
vol. I, ed Greiner W. and Gupta R. K. (Singapore: World
Scientific) pp397, 536.
\bibitem{maruhn74}
Maruhn J and Greiner W, 1974 Phys. Rev. Lett. {\bf 32} 548.
\bibitem{gupta75}
Gupta R. K., Scheid W. and Greiner W., 1975 Phys. Rev. Lett. {\bf 35} 353.
\bibitem{blocki77}
Blocki J. et al, 1977 Ann. Phys. NY {\bf 105} 427.
\bibitem{audi03}
Audi G., Wapstra A. H. and Thibault C., 2003 Nuclear Physics {\bf A 729} p.
337-676.
\bibitem{moller95}
Moller P., Nix J. R., Myers W. D. and Swiatecki W. J., 1995 At. Data Nucl.
Data Tables {\bf 59} 185.
\bibitem{kroger80}
Kr\"oger H. and Scheid W, 1980 J.Phys. G: Nucl. Phys. {\bf 6} L85.
\bibitem{poenaru96}
Poenaru, D. N. and Greiner, W., 1996 editors, Nuclear Decay Modes,
Institute of Physics, Bristol.
\bibitem{poenaru02}
Poenaru D. N. et al., 2002 Phys. Rev. {\bf C 65} 054308.
\bibitem{poenaru06}
Poenaru D. N., Gherghescu R. A. and Greiner W., 2006 Phys. Rev. {\bf C 73} 014608.

\vfill\eject
\newpage
\end{thebibliography}
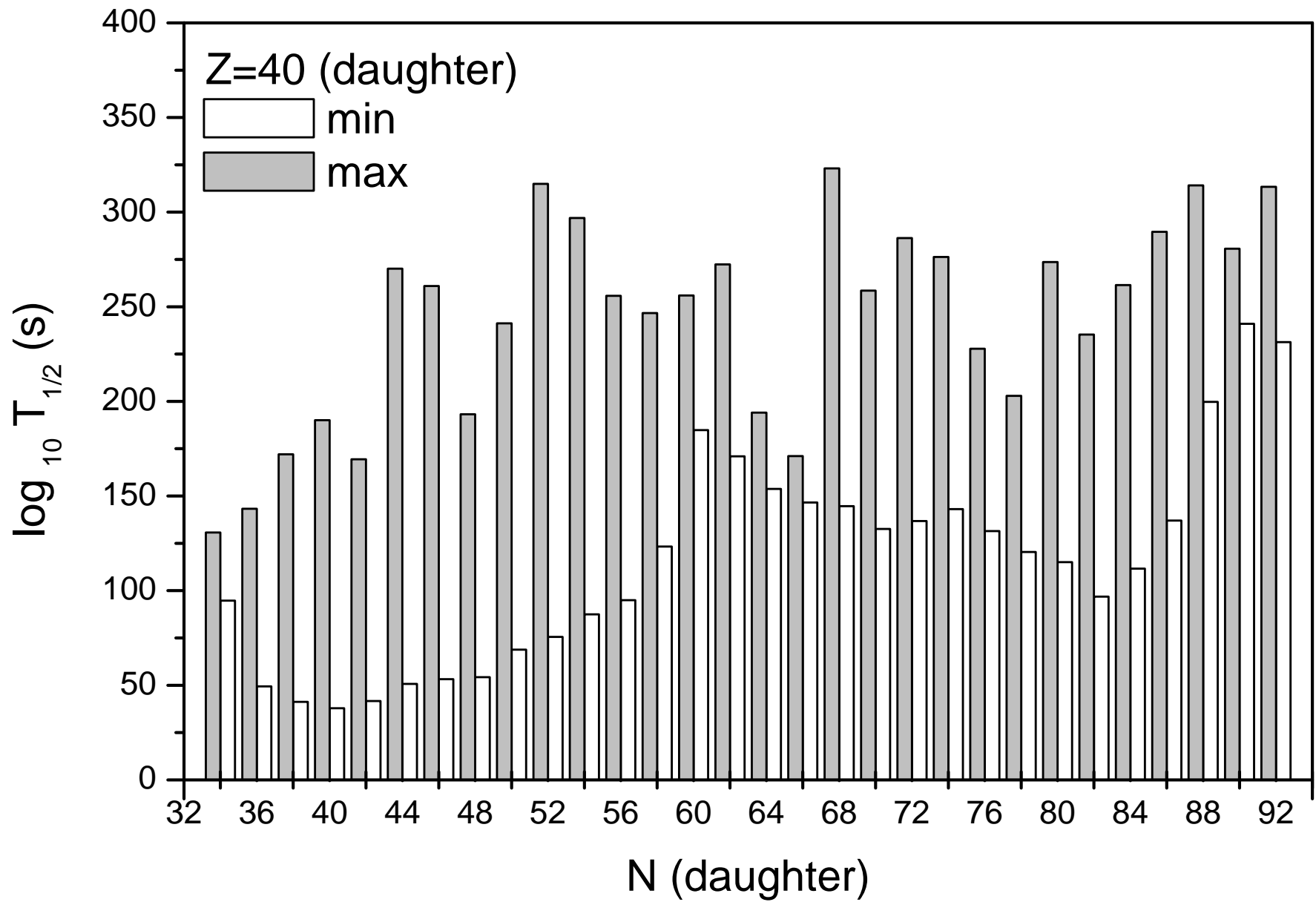
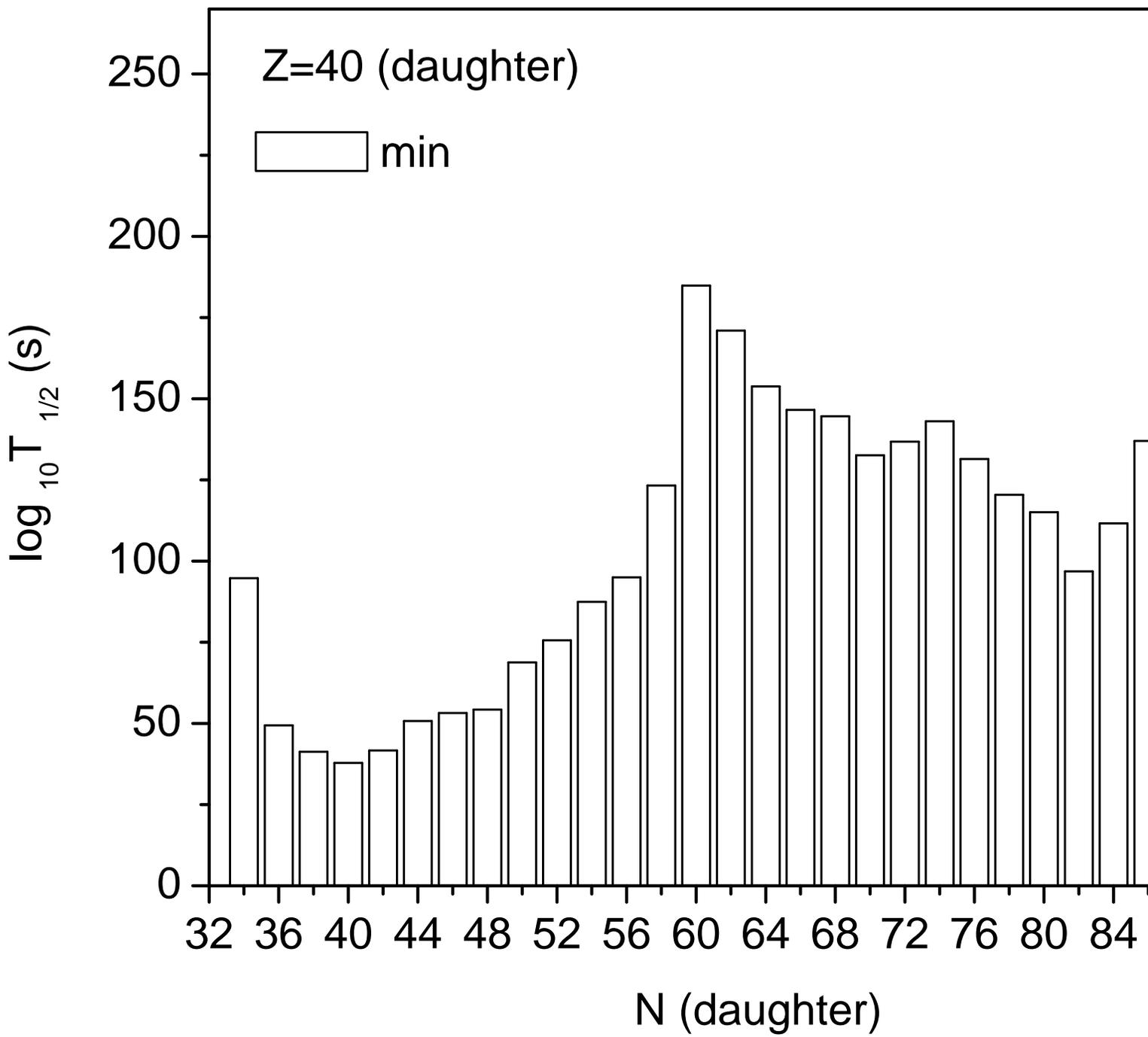
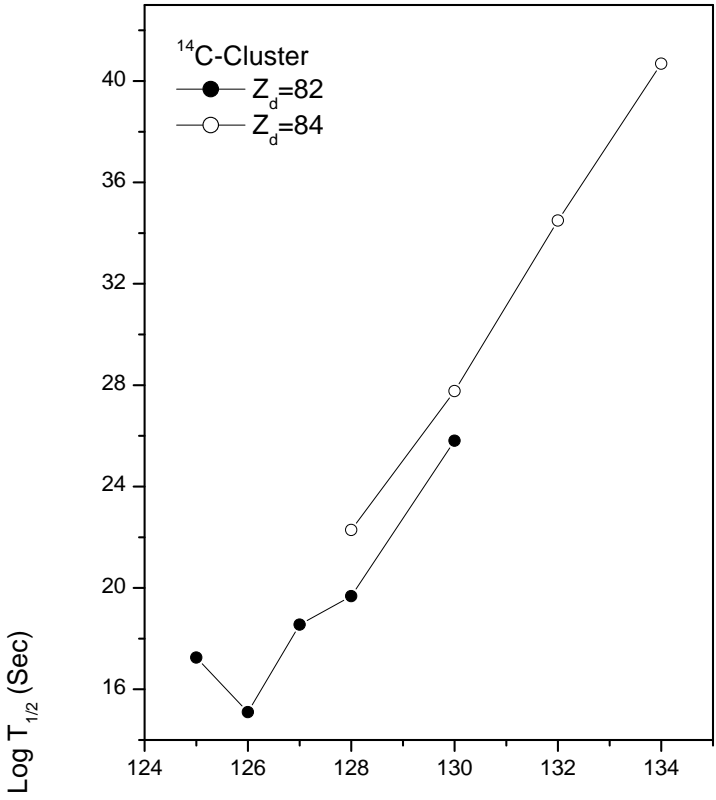
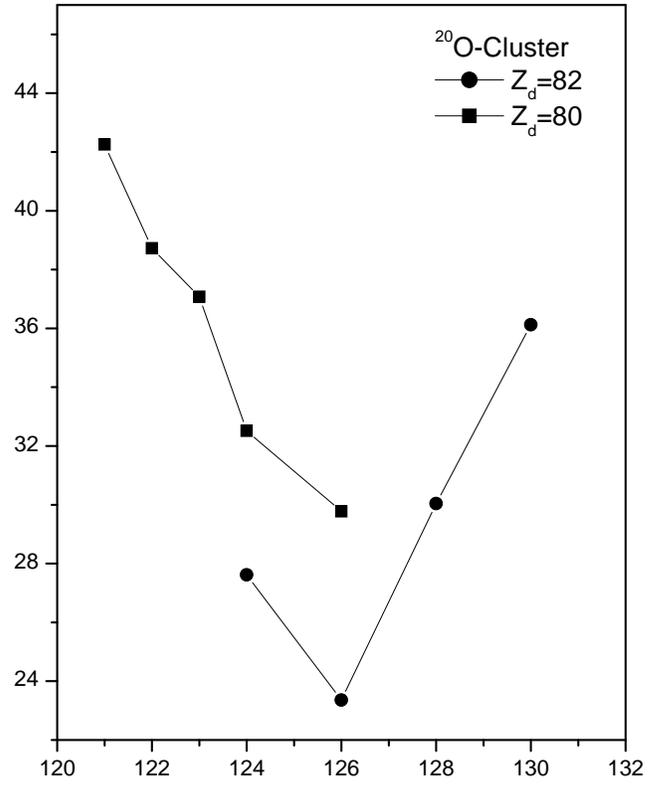
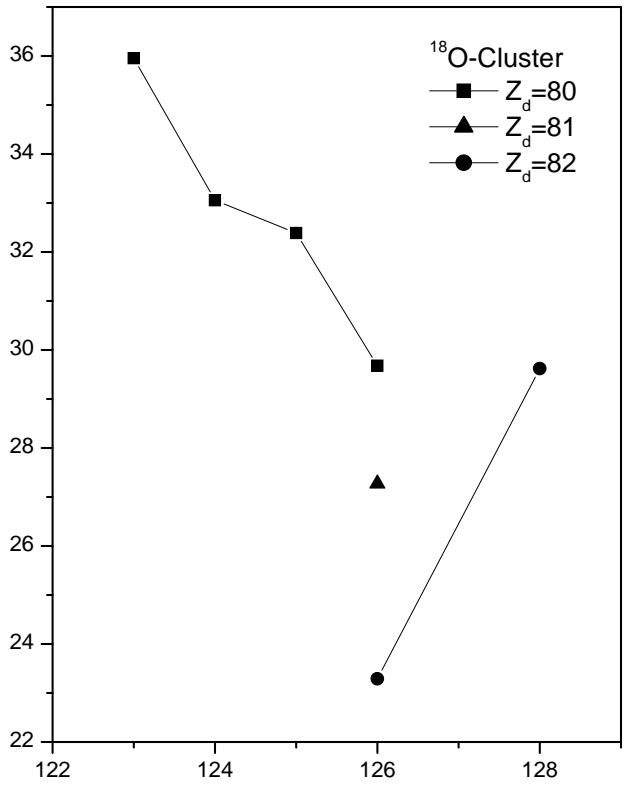
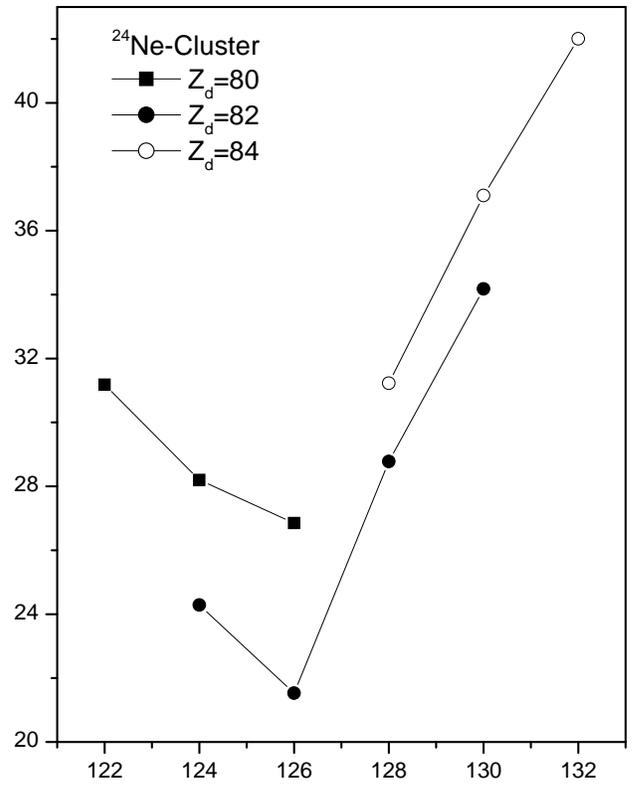
\end{document}